\documentclass[11pt]{article}
\usepackage{times}
\usepackage{mathfont}
\usepackage{graphicx}
\usepackage{cite}

\usepackage{url}
\makeatletter
\def\UrlSpecials{\do\<{\langle}\do\>{\rangle\penalty\relpenalty}\do\_{\_%
 \penalty\@m}\do\|{\mid}\do\{{\lbrace}\do\}{\rbrace\penalty\relpenalty}\do
 \\{\mathbin{\backslash}}\do\~{\mathord{{\textstyle\sim}}}\do\ {\ }}
\makeatother

\def\log{\mathop{\rm log}}

\mathcode`O="724F

\begin{document}

\title{Hinged Kite Mirror Dissection}

\author{David Eppstein\thanks{Dept. Inf. \& Comp. Sci., Univ. of
California, Irvine, CA 92697-3425.  Email: {\tt eppstein@ics.uci.edu}.}}

\date{ }
\maketitle

\begin{abstract}
Any two polygons of equal area can be partitioned into congruent sets of
polygonal pieces, and in many cases one can connect the pieces by
flexible hinges while still allowing the connected set to form both
polygons.  However it is open whether such
a hinged dissection always exists.  We solve a special case of this
problem, by showing that any asymmetric polygon always has a hinged
dissection to its mirror image.  Our dissection forms a chain
of kite-shaped pieces, found by a circle-packing algorithm for
quadrilateral mesh generation. A hinged mirror dissection of a polygon
with $n$ sides can be formed with $O(n)$ kites in $O(n\log n)$ time.
\end{abstract} 

\section{Introduction}

A {\em dissection} of one polygon to another is a partition of the first
polygon into smaller polygonal pieces that can be rearranged to form the
second polygon.  Dissections are possible when (and only when) the two
polygons have the same area, indeed dissection was used by Hilbert as the
basis for an axiomatization of area~\cite{Hil-87}.  Dissection puzzles are
also popular in recreational geometry, where the main aim is to find
dissections of interesting shapes such as regular polygons that use as
few pieces as possible~\cite{Fre-97}.

\begin{figure}[wp]
\centering\includegraphics[width=5.5in]{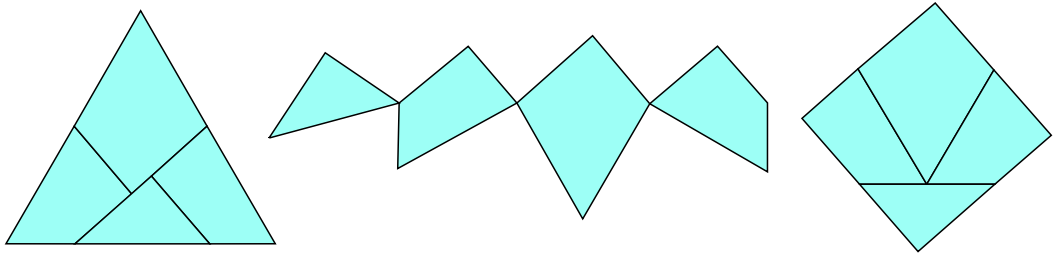}
\caption{Hinged dissection of equilateral triangle and square.}
\label{fig:trisquare}
\end{figure}

\begin{figure}[wp]
\centering\includegraphics[width=3in]{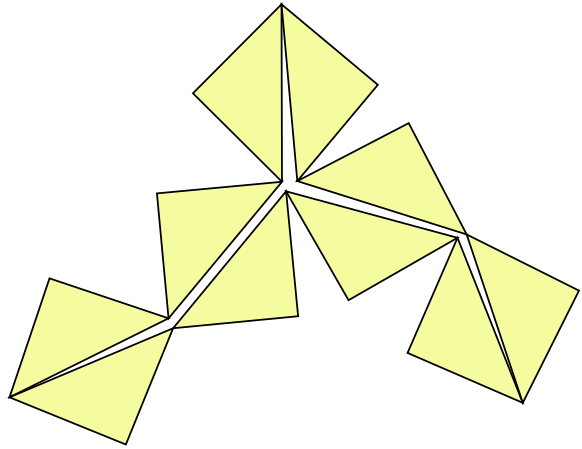}
\caption{Chains of isosceles right triangles form hinged dissections of
any polyomino~\cite{cs.CG/9907018}.}
\label{fig:pentomino}
\end{figure}

\begin{figure}[wp]
\centering\includegraphics[width=\textwidth]{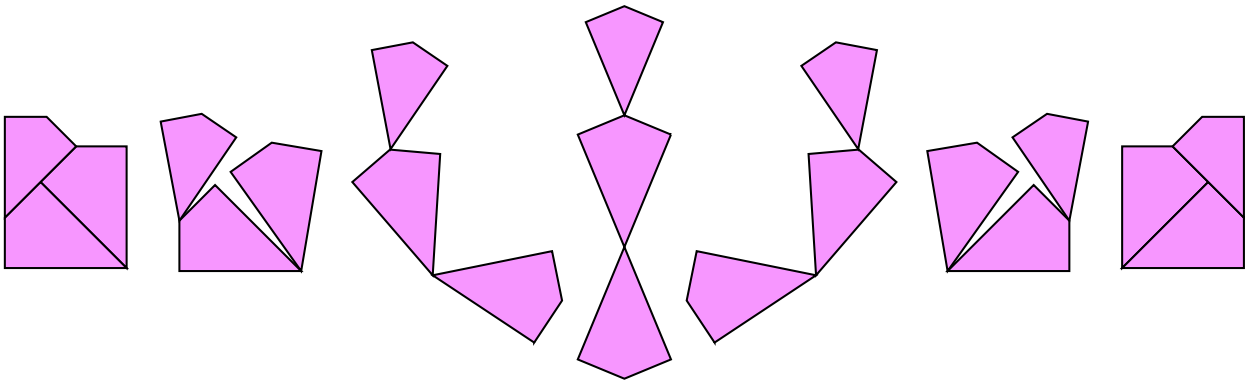}
\caption{A chain of kites, hinged along their axes of symmetry, can be
unfolded so that all axes are colinear, and refolded to form the mirror
image of the original polygon.}
\label{fig:time-lapse}
\end{figure}

A classic example is the four-piece dissection from an equilateral
triangle to a square, often ascribed to Dudeney (but
see~\cite{Fre-97,Fre-02} for speculation on its origin).
The four pieces in the dissection can be connected by {\em hinges},
points of attachment at which the two attached pieces are free to rotate
(Figure~\ref{fig:trisquare}), while still allowing all four pieces to fold
up into both the square and the triangle.  This example has sparked much
interest in similar hinged dissections~\cite{AkiNak-JDCG-98,Fre-02} but
few general results are known, and it remains open whether each equal-area
pair of polygons has a hinged dissection.  In one of the few theoretical
papers in this area, Demaine et al.~\cite{cs.CG/9907018} showed that
chains of isosceles right triangles form hinged dissections between any
pair of $n$-ominos (Figure~\ref{fig:pentomino}), and more generally that
hinged dissections are possibly between many pairs of {\em polyforms},
shapes formed by face-to-face gluing of many copies of the same basic
form.

In this paper we demonstrate the existence of another class of hinged
dissections: we prove that any asymmetric polygon (including polygons with
holes) has a hinged dissection to its mirror image.
Our method is to find a dissection that can be unfolded on its hinges
into a symmetric form: a chain of {\em kites} (quadrilaterals with
reflection symmetry across a diagonal) connected end-to-end, so all
their axes of symmetry lie on a common line.  Clearly,
one can then perform the mirror image of the unfolding process to fold
the chain back up into the mirror image of the original polygon;
Figure~\ref{fig:time-lapse} shows this unfolding and refolding
process applied to a hinged kite dissection of an asymmetric concave
hexagon (however this dissection is not one that would be found by the
algorithms we describe).  We prove that any polygon with
$n$ sides has a hinged kite dissection with $O(n)$ pieces, which can be
computed by an algorithm running in $O(n\log n)$ time.

Although the set of dissections we find is very limited, we discuss
possible implications for more general dissections between any two
equal-area polygons.  Our technique allows us to reduce the general
dissection problem into the question of whether any two equal-area
triangles have a hinged dissection in which some copies of a specified
pair of vertices of the first triangle map to a specified pair of vertices
of the second.

As in the paper of Demaine et al.~\cite{cs.CG/9907018}, we do not
consider the question of whether our dissections can be continuously
unfolded without any intermediate self-intersections, so in the
terminology of Frederickson~\cite{Fre-02} all our dissections are {\em
wobbly-hinged}.

\section{Dissection Process}

\begin{figure}[wp]
\centering\includegraphics[width=5.5in]{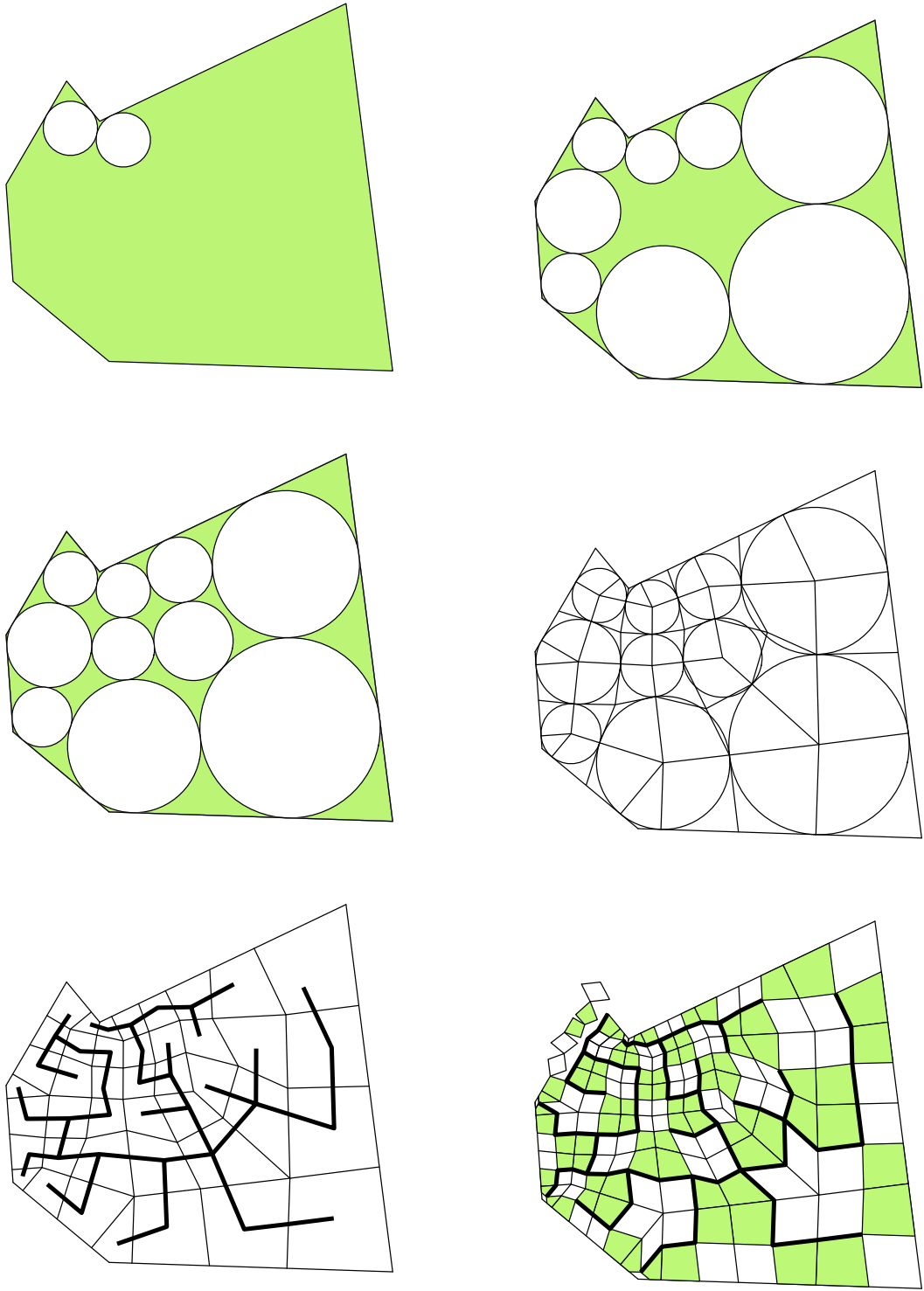}
\caption{Steps in our hinged dissection process.}
\label{fig:kitedis}
\end{figure}

\begin{figure}[t]
\centering\includegraphics[width=4.5in]{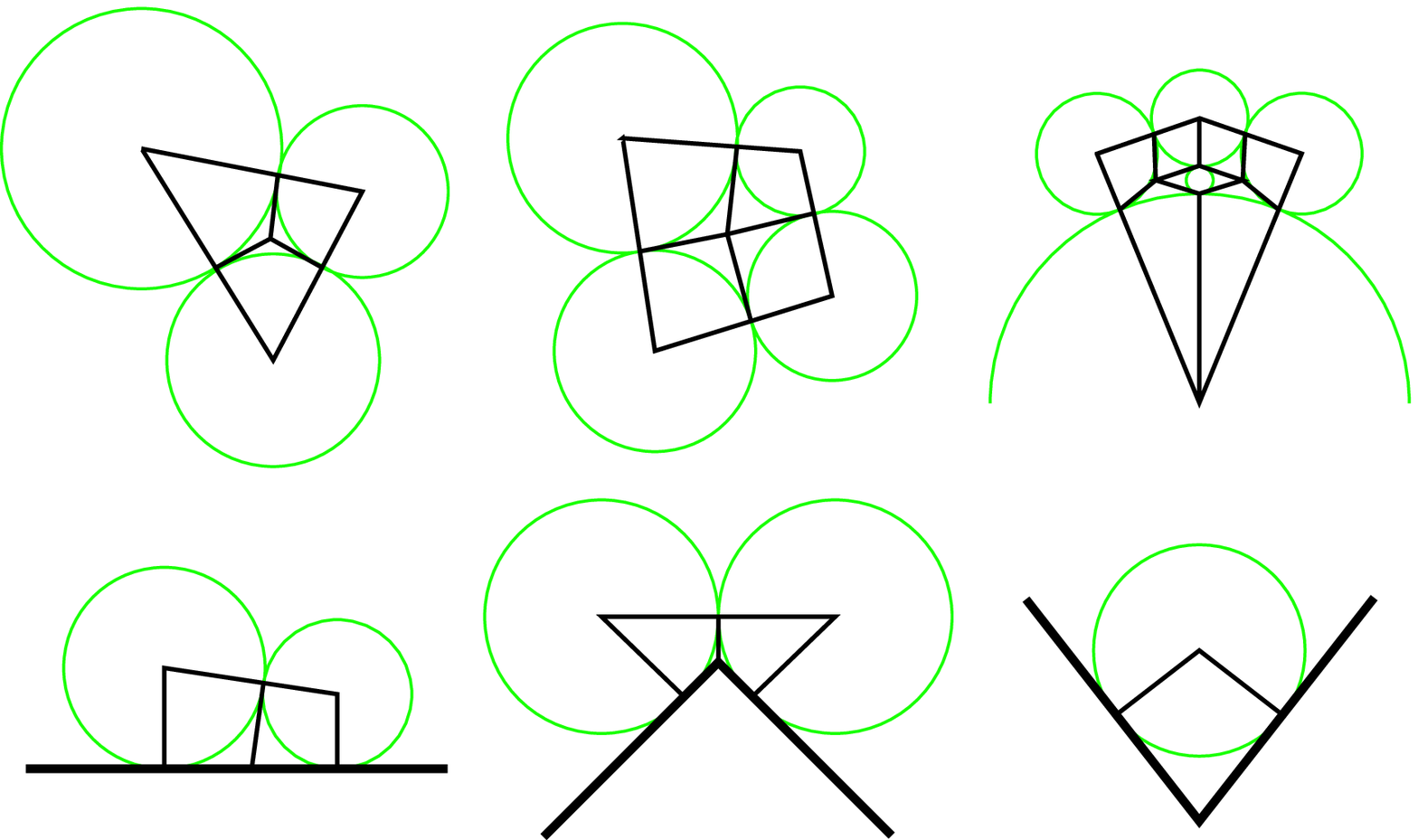}
\caption{Cases for partition of circle-packing gaps into kites,
from~\cite{BerEpp-IJCGA-00}.}
\label{fig:kite-cases}
\end{figure}

We now describe the steps by which we find a hinged dissection of an
arbitrary polygon into a chain of kites. These steps are also illustrated
in Figure~\ref{fig:kitedis}.  Our technique is based on a circle-packing
algorithm of Bern and the author~\cite{BerEpp-IJCGA-00} for partitioning
polygons into well-shaped quadrilaterals in the context of
finite element mesh generation; this method is based on previous
circle-packing nonobtuse triangulation algorithms by Bern et
al.~\cite{BerMitRup-DCG-95,Epp-IJCGA-97} and has also been applied to
problems of paper folding and cutting~\cite{BerDemEpp-Fun-98}.

We dissect the given polygon by the following sequence of steps.
Steps 1-\ref{step-cases} are taken from the kite meshing algorithm of Bern
and the author, while the remaining steps transform the kite mesh into a
hinged dissection.  In steps 1-\ref{finish-packing} we pack the polygon by
tangent circles, so that the polygon is
partitioned by the circles into regions of two types: interiors of
circles, and nonconvex {\em gaps} exterior to the circles.  Each gap is
bounded by three or more {\em sides} consisting of segments of polygon
boundary and arcs of tangent circles.

\begin{enumerate}
\item We begin by placing pairs of congruent circles near each reflex
vertex of the polygon, tangent to each other and to the polygon.

\item We place additional circles tangent to each boundary component of
the polygon, so that the circles are connected in a cycle by
tangencies, with a circle doubly tangent to the polygon near each
convex vertex.  After this step, each gap involving a segment of polygon
boundary (other than the four-sided gaps created in step~1) has exactly
three sides.  However the gaps in the interior of the polygon may still
have many sides.

\item
\label{finish-packing}
As long as there is a gap with five or more sides,
we place a circle to split it into two simpler gaps.

\item
The remaining gaps have three or four sides.  We draw
line segments between each circle center and the circle's points of
tangencies, partitioning the polygon into triangles and quadrilaterals
surrounding each gap, with distinguished points (the tangencies) towards
the center of each triangle or quadrilateral edge.

\item
\label{step-cases}
We now partition each of these triangles or quadrilaterals into kites,
according to a case analysis shown in Figure~\ref{fig:kite-cases}:

\begin{enumerate}
\item In a three-sided gap interior to the polygon, we place a point at
the circumcenter of the triangle formed by the three points of tangency,
and connect this center point to each tangency.

\item In a four-sided gap, the four points of tangency are always
cocircular~\cite{BerMitRup-DCG-95}.  In most such cases, as in the case
of three-sided gaps, we place a point at the circumcenter of these four
points, and connect this center point to each tangency.

\item There may be some four-sided gaps in which the center point is not
interior to the convex hull of the four tangencies, so that the previous
case would lead to the creation of a concave dart shape instead of a
kite.  Bern et al.~\cite{BerMitRup-DCG-95} call this case a {\em bad gap}
and show that it can always be split into two good gaps by the addition
of a single circle tangent to two of the four arcs of the bad gap.
These two good gaps can be covered by seven kites
(Figure~\ref{fig:kite-cases}, top right).

\item When two circles form a gap with a straight piece of polygon
boundary, we can partition this gap into two kites by a line segment
through the circle tangency and perpendicular to the line between the two
circle centers.  The same type of partition also applies to a gap
containing a reflex vertex, because we chose the two circles
forming this gap to be congruent.

\item The final case consists of a gap formed by a convex vertex and a
single circle, however this type of gap is already in the form of a kite.
\end{enumerate}

\item
\label{tree-step}
We now have a partition of the polygon into kites,
however we are not finished because it may not be possible to hinge the
kites appropriately.  We call the kites of this partition {\em large
kites} to distinguish them from the ones formed in step~\ref{step-bisect}
below. We next find a tree, with one vertex interior to each large kite,
where each tree edge connects points from two adjacent large kites. 
(I.e., this is a spanning tree of the dual graph of the large kite mesh.)

\item
\label{step-bisect}
We partition each large kite into four smaller pieces by
placing a point at the intersection of its two diagonals, and connecting
that point to the midpoints of the large kite edges.  This partitions the
large kite into four pieces, two of which are similar to the original
large kite (shown shaded in Figure~\ref{fig:kitedis}) and the other two of
which are rhombi.

\item We arrange the spanning tree of step~\ref{tree-step} so that its
vertices lie at the interior points added within each large kite, and its
edges lie along the connections from these interior points to the
large kite edge midpoints.  We add a single segment connecting this
spanning tree to the midpoint of an edge on the outer boundary of the
polygon.

\item Finally, we trace around the boundary of the tree, and form a
linear sequence of small kites and rhombs in the order in which they are
visited by this trace.  We hinge these kites and rhombs at the vertices
on the edge midpoints of the large kites.  Each small kite is hinged at
the two vertices of its axis of symmetry, and each rhomb is hinged at
two opposite vertices. 
\end{enumerate}

\begin{figure}[t]
\centering\includegraphics[width=5.5in]{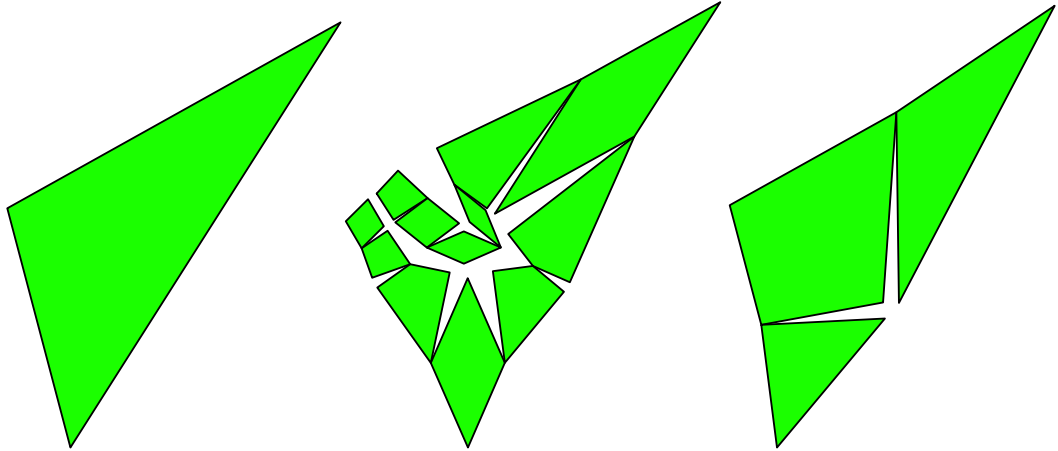}
\caption{Hinged dissections of a scalene triangle.}
\label{fig:mirror-tri}
\end{figure}

Bern and the
author~\cite{BerEpp-IJCGA-00} use a somewhat more complicated case
analysis in step~\ref{step-cases}, allowing four-sided gaps involving the
edges of the polygon, in order to show that the dissection into kites
used here can be performed in time
$O(n\log n)$ and that it need only create
$O(n)$ pieces.

Another example of a dissection created by this process, of a scalene
triangle, is shown in Figure~\ref{fig:mirror-tri} (center).  In this case
the circle packing consists of a single circle inscribed in the triangle,
eventually resulting in a twelve-piece dissection. However, the number of
pieces can be improved: as shown in the right of the figure, any scalene
triangle has a three-piece hinged mirror dissection into a kite and two
isosceles triangles, formed by cutting from the midpoints of the two
short sides of the triangle to a third point on the hypotenuse. The third
point is the reflection of the hypotenuse midpoint across the
perpendicular bisector of the two other midpoints; the line through it
and the opposite vertex is perpendicular to the hypotenuse. The three
pieces formed by these two cuts are then hinged at the side midpoints.

\section{Possible Implications}

We still seem to be a long way from solving the question of whether
hinged dissections exist between any pair of equal area
polygons, or even more generally between any set of equal area polygons.
However, our kite dissection can be used to reduce this problem to a
seemingly more simple form.

\begin{figure}
\centering\includegraphics[width=3.5in]{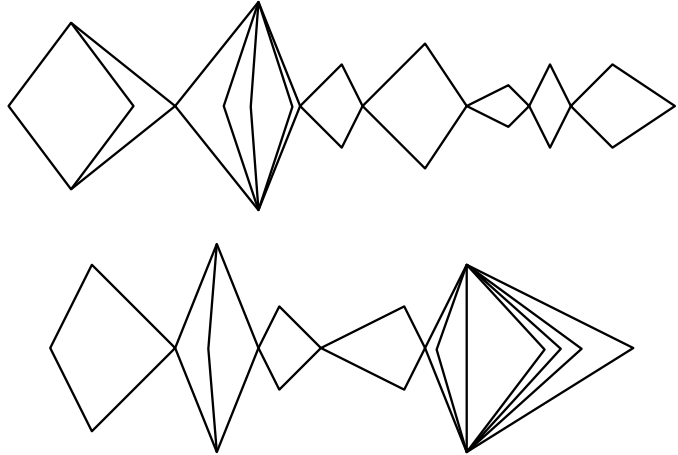}
\caption{Refinement of two equal-area chains of kites into smaller kites
and darts so that each pair of pieces has the same area.}
\label{fig:kitesplit}
\end{figure}

Suppose we have two equal area polygons, both dissected into chains of
kites hinged end-to-end.  The sequences of areas of the kites can be
viewed as partitions of the one-dimensional interval $[0,A]$, and we can
find a common refinement of these two partitions by overlaying them.
Geometrically, as shown in Figure~\ref{fig:kitesplit}, this corresponds
to introducing a sequence of cuts to the two chains of kites,
partitioning them into smaller kites and darts, still hinged end-to-end,
so that both chains are composed of polygons that form the same sequence
of areas. In other words, the first kite or dart from the first chain has
the same area as the first kite or dart from the second chain, and so on.
The cuts in one chain correspond to the hinges in the other chain and
vice versa.  If we could then further hinge-dissect each equal-area pair,
we could combine these parts into a hinge dissection of the original two
polygons. By further splitting the kites and darts along their axes of
symmetry, we reduce the problem to one in which we must dissect a
sequence of equal area triangles.

Summarizing, we would be able to hinge-dissect any two equal-area
polygons, if only we could hinge-dissect the very simple special case of
two equal-area triangles, with the restriction that copies of two
vertices from the first triangle are mapped to two vertices of the second
triangle so that the dissected triangles can be connected to their
neighbors in the chain.  Our kite dissection method transforms any single
dissection problem of two polygons with a total of $n$ sides into a
sequence of
$O(n)$ triangle dissection problems.  More generally, we can use the same
construction to reduce any $k$-way dissection problem to one involving
only triangles.

A similar result could be obtained by using Saalfeld's decomposition of
equal-area polygons into combinatorially equivalent equal-area
triangulations~\cite{Saa-SCG-01}, however his method lacks complexity
bounds and seems to use a large number of pieces.

\section{Discussion}

We have shown that any polygon has a hinged dissection in
the form of a chain of kites, that can be unfolded and refolded to
form the mirror image of the original polygon.  The result also has some
possible consequences for the open problem of the existence of hinged
dissections between any pair of equal area polygons.

Some questions about our method remain unanswered, for instance whether
our dissections or modifications of them can be unfolded in a continuous
motion that avoids self-intersections.  Also, the number of pieces used
by our dissections, although asymptotically optimal, seems large, and
Figure~\ref{fig:mirror-tri} shows that it can be reduced by a factor of
four in the case of scalene triangles.  Is a similar reduction possible
more generally?

\section*{Acknowledgements}

My thanks go to Greg Frederickson for encouraging me to publish these
results, to Erik Demaine for extensive comments on a draft of this paper,
and to Cinderella for help with the figures.
This work was supported in part by NSF grant CCR-9912338.

\bibliographystyle{abuser}
\bibliography{cpack}

\end{document}